\documentclass[superscriptaddress,twocolumn,showpacs,pra,nofootinbib]{revtex4-2}
\usepackage{bm} 
\usepackage{graphicx} 
\usepackage{amsmath}
\usepackage{amsthm}
\usepackage{amssymb} 
\usepackage{epstopdf} 
\usepackage{comment}
\usepackage{amsfonts}
\usepackage{epsfig}
\usepackage{dsfont}
\usepackage{color} 
\usepackage[english]{babel}
\usepackage[bb=boondox]{mathalfa}
\usepackage{mathtools}
\usepackage{hyperref}
\hypersetup{
    colorlinks=true,       
    linkcolor=cyan,          
    citecolor=magenta,        
    filecolor=magenta,      
    urlcolor=cyan,           
    runcolor=cyan
}

\newcommand{\bra}[1]{\langle #1 |}
\newcommand{\ket}[1]{| #1 \rangle}

\newcommand {\be}{\begin{equation}}
\newcommand {\ee}{\end{equation}}

\newcommand\ua{\uparrow}
\newcommand\da{\downarrow}

\newcommand\zl{\langle z_{\textrm{out}}|}
\newcommand\zr{|z_{\textrm{in}}\rangle}

\newcommand\zinout{\textrm{in}\to \textrm{out}}

\newcommand\me{{\cal A}_{\zinout}}

\newcommand{\ba}{\begin{eqnarray}}
\newcommand{\ea}{\end{eqnarray}}

\newcommand\tr{{\mbox{Tr\,}}}
\newcommand{\ignore}[1]{}

\usepackage[margin=1in,nofoot]{geometry}

\newcommand{\Tr}{{\mathrm{Tr}}}
\newcommand{\e}{{{e}}}
\newcommand{\rmd}{{\text d}}

\newcommand{\beq}{\begin{equation}}
\newcommand{\eeq}{\end{equation}}
\newcommand{\beqnn}{\begin{equation*}}
\newcommand{\eeqnn}{\end{equation*}}
\newcommand{\bea}{\begin{eqnarray}}
\newcommand{\eea}{\end{eqnarray}}
\newcommand{\beann}{\begin{eqnarray*}}
\newcommand{\eeann}{\end{eqnarray*}}
\newcommand{\bes} {\begin{subequations}}
\newcommand{\ees} {\end{subequations}}

\begin{document}

\title{Feynman path integrals for discrete-variable systems: \\
Walks on Hamiltonian graphs}
\author{Amir Kalev}
\affiliation{Information Sciences Institute, University of Southern California, Arlington, VA 22203, USA}
\affiliation{Department of Physics and Astronomy, and Center for Quantum Information Science \& Technology,University of Southern California, Los Angeles, California 90089, USA}
\author{Itay Hen}
\affiliation{Information Sciences Institute, University of Southern California, Marina del Rey, CA 90292, USA}
\affiliation{Department of Physics and Astronomy, and Center for Quantum Information Science \& Technology,University of Southern California, Los Angeles, California 90089, USA}

\begin{abstract}
\noindent We propose a natural, parameter-free, discrete-variable formulation of Feynman path integrals. 
We show that for discrete-variable quantum systems, Feynman path integrals take the form of walks on the graph whose weighted adjacency matrix is the Hamiltonian. By working out expressions for the partition function and transition amplitudes of discretized versions of continuous-variable quantum systems, and then taking the continuum limit,  we explicitly recover Feynman's continuous-variable path integrals. We also discuss the implications of our result.
\end{abstract}

\maketitle

\section{Introduction}

Feynman's path integral formulation of quantum mechanics~\cite{feynman1,feynman2} is rightfully considered a cornerstone of modern theoretical physics.  While equivalent to the Schr{\"o}dinger formulation, the path integral view has introduced a novel way of understanding quantum phenomena and as such it has had a profound impact on theoretical physics, providing a deeper understanding of quantum mechanics, enabling calculations in fields ranging from particle physics to condensed matter physics~\cite{PIbook,CMTbook}. The importance of this formulation lies in several key aspects. First and foremost, Feynman's approach provides a more intuitive understanding of quantum mechanics compared to traditional formulations. Rather than focusing on wave functions and operators, it emphasizes the concept of paths. Each possible path a particle can take contributes to the overall probability amplitude, and the interference of these paths determines the behavior of quantum systems. This intuitive picture has proved invaluable in teaching and conceptualizing quantum mechanics.

In the limit of large action (equivalently, small Planck's constant), the path integral formulation recovers classical physics through the principle of least action providing a deeper insight into the classical limit of quantum systems. Feynman's path integral formulation also played a crucial role in the development of quantum field theory, providing a framework for calculating scattering amplitudes and understanding the behavior of quantum fields in terms of particle interactions. 

While the path integral formulation is extremely powerful and can be applied to a wide range of physical systems, including ones with complex interactions and boundary conditions,
the discrete-variable analog of path integrals, namely a path integral formulation for finite-dimensional or discrete quantum systems, has long been a rather elusive concept -- a limitation that was recognized already by Feynman~\cite{feynman1}. The main reason for the above deficiency is that the Feynman path integral formulation relies on the existence of a {\it classical} description of the system, namely on the system Lagrangian, which most discrete-variables quantum systems, most notably spin systems, do not naturally have. Over the years multiple attempts to close this gap have been made (see, e.g., Refs~\cite{karchev2012path,PhysRevResearch.5.043075}) with the main idea being finding continuous classical analogs for spin operators~\cite{CMTbook}. Examples include the use of spin coherent states~\cite{klauder,JOLICOEUR2019445,garg2001spincoherentstatepathintegrals} or discrete time representations in combination with Grassmann variables non-orthogonal coherent states~\cite{Grinberg2003PathIO} for specific models. These too however have mostly been found to be either model specific, impractical or riddled with problems~\cite{doi:10.1142/S0217979299000096}. Other approaches introducing rather general procedures for constructing Feynman path integrals directly from the Hamiltonian have also been proposed. One example  is the construction proposed by Farhi and Gutmann~\cite{FARHI1992182} where one is required to consider all mappings between continuous time intervals and sequences of off-diagonal Hamiltonian matrix elements. Also noteworthy in that regard is the seminal work by Prokof’ev and Svistunov~\cite{qpt:book,PhysRevE.74.036701,PROKOFEV1998253} which gave birth to continuous-time path integral quantum Monte Carlo. There, discrete-variable paths are obtained from sampling the integrands of the Dyson series expansion of the Hamiltonian exponential. 

Nonetheless, an all-encompassing canonical formulation of Feynman path integrals applicable to finite-dimensional systems as a whole has remained missing.

Here we propose an altogether different natural formulation of Feynman path integrals for discrete quantum systems. In our approach, which does not rely on the existence of a classical description of the system, the analogs of Feynman paths are walks on a graph induced by the Hamiltonian of the system, and path integrals take the form of sums over all relevant walks on said graph. 
We explicitly show that our {\it walk sum} formulation is capable of recovering the appropriate Feynman path integrals in the continuum limit. 

The main workhorse in our derivation is the off-diagonal series expansion for matrix functions 
-- a power-series expansion for matrix functions in the off-diagonal strength of the input matrix -- 
recently introduced in the context of quantum Monte Carlo simulations~\cite{ODE,pmr,fmat}. Under this expansion, matrix elements of matrix functions are represented as sums over terms each of which corresponding to a walk on the Hamiltonian graph -- the graph for which the Hamiltonian serves as a weighted adjacency matrix~\cite{PhysRevResearch.3.023080}. 

In what follows, we first derive the discrete version of path integrals and in the sections that follow, we analyze it for a discretized continuous-variable one-dimensional quantum system and show that in the continuum limit, the it recovers Feynman path integral of the partition function and transition amplitude. We also analyze in detail a discrete-variable calculation choosing as our test model the quantum transverse-field Ising model and illustrate how transition amplitude calculations for this model are carried out. Finally, we offer conclusions and discuss the implications of our result. 

\section{The off-diagonal series expansion and walk sums}
To arrive at a discrete-variable formulation of Feynman path integrals, 
we consider a finite-dimensional (or countable infinite-dimensional) Hamiltonian, $H$ and write it in permutation matrix representation (PMR) form~\cite{ODE, ODE2, pmr}:
\beq \label{eq:basic}
H=\sum_{j=0}^M D_j P_j   = D_0 + \sum_{j=1}^M D_j P_j \,. 
\eeq
In PMR, the Hamiltonian is written as a sum of well-defined operators represented in some basis ${\cal{B}}$ whose basis states we denote $\{\ket{z}\}$. Each operator $D_j$ (with $j=0,\ldots,M$) is a diagonal operator in ${\cal{B}}$ while each $P_j$ is a distinct  \emph{permutation} operator~\cite{gpm} whose  matrix representation in the ${\cal{B}}$ basis  has exactly one entry of 1 in each row and each column and all other elements are zero. One may always choose the matrices $P_j$ such that none of them, nor any product thereof, have fixed points (that is, no nonzero diagonal elements) with the exception of the identity matrix $P_0=\mathds{1}$. The diagonal matrix $D_0$ is the diagonal component of $H$. Furthermore, each term $D_j P_j $ obeys 
$D_j P_j |z\rangle = d_j(z') | z' \rangle$ where $d_j(z')$ is a possibly complex-valued coefficient and $|z'\rangle \neq |z\rangle$ is a basis state. Casting Hamiltonians in PMR form can always be done efficiently~\cite{pmr}. 

Next, consider an arbitrary Taylor-expandable function of $H$, namely $f(H)$, 
\bea
f(H)&=& \sum_{n=0}^{\infty}\frac{f^{(n)}(0)}{n!} H^n\nonumber\\&=&\sum_{n=0}^{\infty}\frac{f^{(n)}(0)}{n!}   (D_0+\sum_j D_j P_j)^n\nonumber\\&=& \sum_{n=0}^{\infty}  \sum_{{\bf i}_n} \frac{f^{(n)}(0)}{n!}   S_{{{\bf i}}_n}  \,,
\eea
where $f^{(n)}(0)$ is the $n$-th derivative of $f(\cdot)$ at zero. In the above expression, $S_{{{\bf i}}_n}$ denotes all operator sequences of length $n$ consisting of (ordered) products of basic operators $D_0$ and $D_j P_j$. The index ${\bf i}_n= (i_1,\ldots,i_n)$ is a multiple index where each individual index $i_j$ (with $j=1\ldots n$) ranges from $0$ to $M$. In Refs.~\cite{ODE,ODE2,pmr,fmat} it was shown that the matrix element of  $f(H)$ between two (basis) states $|z_{\textrm{in}}\rangle$  and $|z_{\textrm{out}}\rangle$ is given by 
\begin{align} \label{eq:z1}
&\zl f(H) \zr\\\nonumber&= \sum_{q=0}^{\infty}  \sum_{{{\bf i}}_q}  D_{{\bf i}_q}  {f [E_{z_0},\ldots,E_{z_q}]} \zl P_{{\bf i}_q} \zr \,,
\end{align} 
where $P_{{\bf i}_q}=P_{i_q}\cdots P_{i_1}$ denotes a sequence of length $q$ permutation operators, $E_{z_i}=\langle z_i |D_0 | z_i \rangle$ (where $\ket{z_0}\equiv\zr$), and 
\begin{align}
&D_{{\bf i}_q}=\prod_{j=1}^q d^{(i_j)}_{z_j},\text{ where }   d^{(i_j)}_{z_j} = \langle z_j|  D_{i_j}|z_j\rangle,\\\nonumber &\text{with } |z_j\rangle=P_{i_j} \ldots P_{i_2} P_{i_1} \zr.
\end{align}
Note that $|z_j\rangle$ should have been denoted  in principle  by $|z_{{\bf i}_j}\rangle$ however we are using a simplified notation so as not to overburden the notation.
The coefficients $d^{(i_j)}_{z_j}$ may be viewed as the  `hopping strengths' of $P_{i_j}$ with respect to $|z_j\rangle$, and the sequence of basis states $\{|z_j\rangle \}$ may be viewed as a `walk' in the graph of basis states induced by the Hamiltonian~\cite{ODE,ODE2,signProbODE}. In addition, the term ${f[E_{z_0},\ldots,E_{z_q}]}$ is the \emph{divided differences} of $f(\cdot)$ over the multi-set of diagonal element (energies) $[E_{z_0},\ldots E_{z_q}]$~\cite{dd:67,deboor:05,pmr}.

\begin{figure}[htp!]
\begin{center}
\includegraphics[width=0.98\columnwidth]{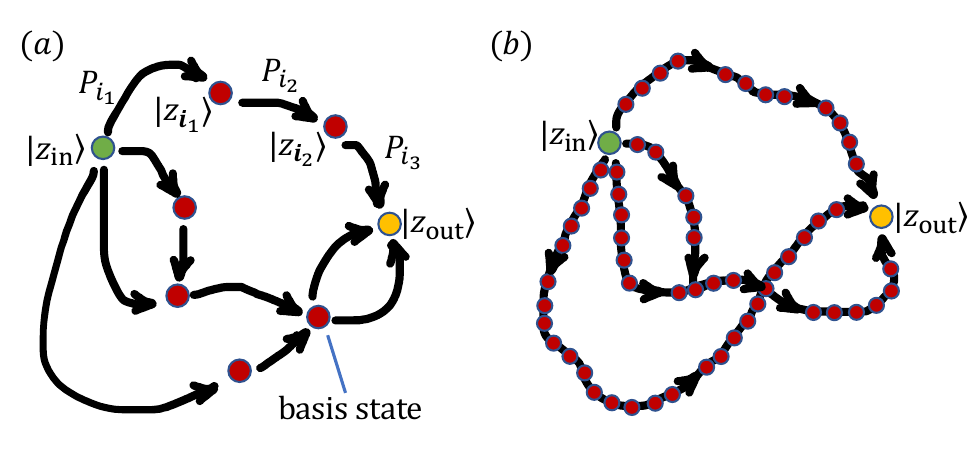}
\end{center}
\caption{{\bf (a) Matrix elements as walks over the Hamiltonian graph}. A matrix element $\zl f(H) \zr$ is calculated by summing up divided-difference contributions from all walks leading from $\zr$ to $|z_{\textrm{out}}\rangle$ along the edges of the Hamiltonian graph. Here, each node represents a basis state and each (directed) edge -- an off-diagonal Hamiltonian matrix element. {\bf (b) Walks on the Hamiltonian graph as the continuum limit is approached.} In the continuum limit, walks become Feynman paths.}
\label{fig:walks}
\end{figure}

Since $P_j$ for all $j$ is a permutation operator with no fixed point, $P_{{\bf i}_q}$ is a permutation operator, and therefore the expression $\zl P_{{\bf i}_q} \zr$ evaluates either to 1 or to zero. Thus, we can recast Eq.~\eqref{eq:z1} as
\begin{align} \label{eq:z2}
\zl f(H) \zr= \sum_{q=0}^{\infty}  \sum_{{{\bf i}}_q : \zinout} \!\!\! D_{{\bf i}_q}  {f [E_{z_0},\ldots,E_{z_q}]} \,.
\end{align} 
where in the summation above, ${\bf i}_q : \zinout$, is shorthand for a sum over all sequences (or walks) ${\bf i}_q$ such that $\zl P_{{\bf i}_q} \zr=1$. Equation~\eqref{eq:z2} captures the mathematical essence of the PMR formalism. It implies that any matrix element of a function of the Hamiltonian, $\zl f(H) \zr$, can be written as a sum  over all possible walks on the Hamiltonian graph that 
connect $\ket{z_{\rm in}}$ with $\ket{z_{\rm out}}$. The weight of each walk is given by a product of hop strengths, $D_{{\bf i}_q}$, and the corresponding divided differences of $f(\cdot)$ with `classical' energy inputs. This result is illustrated in Fig.~\ref{fig:walks}(a). 

We next argue that the above representation is a natural discrete-variable of Feynman path integral, wherein matrix elements are represented as a sum over terms corresponding to all viable paths (more precisely, walks) between the appropriate endpoints. Let us now write an explicit `walk sum' expression for the canonical partition functions and for transition amplitudes under time evolution. For the canonical partition function, the walk sum is obtained by taking $f(\cdot)=\e^{-\beta (\cdot)}$ summing together all closed walks, that start and end at the same state to encapsulate the trace operation. Equation~\eqref{eq:z2} reduces in this case to
\begin{align} \label{eq:z2Z}
Z=\sum_z\bra{z}\e^{-\beta H}\ket{z}=\sum_{z,q} \sum_{{\bf i}_q:z\to z} D_{{\bf i}_q}   \e^{-\beta [E_{z_0},\ldots,E_{z_q}]}\,,
\end{align}
where the  sum  over ${\bf i}_q:z\to z$ above is a sum over  all closed walks (to be compared against Feynman's closed paths in the case of partition function representation). For quantum transition amplitudes, the walk sum is obtained by taking $f(\cdot)=\e^{-\frac{i}{\hbar} t(\cdot)}$, in which case the matrix function becomes a quantum time-evolution operator with $t$ representing time (we assume for simplicity that the Hamiltonian is time independent; a generalization of the off-diagonal formulation to the time-dependent regime also exists~\cite{Kalev_2021}). In this case, the matrix element expresses the transition amplitude from $\zr$ to $|z_{\textrm{out}}\rangle$ after time $t$ under the Hamiltonian $H$, explicitly: 
\begin{align}\label{eq:z2A}
\me &=\zl\e^{-\frac{i}{\hbar} t H} \zr\\\nonumber&=\sum_{{\bf i}_q : \zinout} D_{{\bf i}_q}   \e^{-\frac{i}{\hbar} t [E_{z_0},\ldots,E_{z_q}]}\,.
\end{align}
We thus find that  transition amplitudes   may be computed by summing divided-difference `phases' for all walks on the Hamiltonian graph between the two states (this is to be compared against Feynman path integral which prescribe summing over the phases of all paths connecting the two states). 

We next show how the familiar continuous-variable Feynman path integral formulation is naturally recovered in the continuum limit, from the above sum over walks on the Hamiltonian graph.

\section{The Feynman path integral as the continuum limit of the walk sum formulation}

Consider a single particle of mass $m$ moving in one dimension under the influence of an arbitrary potential (it should be clear that the extension to multiple dimensions and more complex scenarios can be derived in the exact same manner).
The Hamiltonian of the system is given by
\beq\label{eq:cont ham}
H=\frac{\hat{p}^2}{2 m} +V(\hat{x}) =-\frac{\hbar^2}{2m} \frac{\partial^2}{\partial x^2} + V(\hat{x}) \,,
\eeq
where $\hat{x}$ and $\hat{p}$ are the position and momentum operators of the particle, respectively.
Using the Feynman path integral formulation, the canonical partition function for the above Hamiltonian can be written as~\cite{PIbook} 
 \beq\label{eq:FeyZ}
Z=\oint {\cal D} x(t) \e^{-\int_0^{\beta} \rmd t \left( \frac{m}{2}\dot{x}(t)^2 + V(x(t))\right) } \,,
\eeq
where $\oint {\cal D}x(t)$ stands for an integral over all closed continuous paths on the real line obeying $x(0)=x(\beta)$. Similarly, the transition amplitude between to arbitrary endpoints $x(0)$ and $x(t)$ is given
by
\beq\label{eq:FeyA}
{\cal A}_{{\rm in}\to{\rm out}}=\int_{x(0)=x_{\textrm{in}}}^{x(t)=x_{\textrm{out}}}  {\cal D}x(t) \e^{i\int_0^t \rmd \tau \left(\frac{m}{2 } \dot{x}(\tau)^2 - V(x(\tau))\right)} \,.
\eeq

To recover the above expressions from their discrete-variable counterparts in the continuum limit, let us discretize the system Hamiltonian in the standard manner.  We replace the real-numbers line with a discrete set of equally spaced points, with some spacing $a$.  The position states now become countable. We denote this set of states by $\{|j\rangle\}_{-\infty}^{\infty}$ where $|j\rangle$ corresponds to the location $x_j=j a$. 
The discretized version of the Hamiltonian, Eq.~\eqref{eq:cont ham}, can therefore be written
 as~\cite{Boykin2004,TARASOV201668} (we fix $\hbar=1$ for convenience): 
\begin{align}\label{eq:dis ham}
H_a&=-\frac{1}{2m a^2}\sum_{j=-\infty}^\infty \left( |j+1\rangle\langle j| -2 |j\rangle\langle j| +|j-1\rangle\langle j|\right) \nonumber\\&+\sum_{j=-\infty}^\infty V(j a)|j\rangle \langle j|  \,.
\end{align}
In PMR form, $H_a$ is given by 
\beq\label{eq:dis ham 2}
H_a = -\frac{1}{2 m a^2} \left( P_{+} + P_{-}\right) +\sum_{j=-\infty}^\infty E_j^{(a)} |j\rangle \langle j| \,,
\eeq
where $E_j^{(a)} = V(j a)+\frac{1}{m a^2}$ and \hbox{$P_{\pm} = \sum_{j=-\infty}^\infty |j \pm 1\rangle\langle j|$} are permutation operators that shift the position of the particle by one spacing in $\pm$ direction. 

Following the derivations of the previous section, the `walk sum' of the partition function corresponding to $H_a$ is given by
\begin{align}
Z_a&=\Tr[\e^{-\beta H_a}]\\\nonumber&=\sum_{j,q}\sum_{{\bf i}_q: j\to j}\left(-\frac{1}{2 m a^2}\right)^q  
\e^{-\beta[E^{(a)}_0,\ldots,E^{(a)}_q]}\\\nonumber&=\sum_{j,q}\sum_{{\bf i}_q: j\to j}\left(-\frac{1}{2 m a^2}\right)^q \e^{-\frac{\beta}{m a^2}} 
\e^{-\beta[V(j_0 a),\ldots,V(j_q a)]},
\end{align}
where in the last step we used the divided-differences property~\cite{ODE,pmr} $e^{[E_0+\Delta,\ldots,E_q+\Delta]}=e^{\Delta}e^{[E_0,\ldots,E_q]}$, and ${\bf i}_q=(i_1,\ldots, i_q)$ with $i_k=\pm, k\in\{1,\ldots,q\}$, $j_k=j_0+\sum_{\ell=1}^k i_\ell$, and where for convenience we have defined $j_0 \equiv j$. We further note that the summation over all closed-loop paths $\sum_{{\bf i}_q: j\to j} $ is equivalent to requiring that $j_q=j_0$, and therefore we can write
\begin{align}
Z_a=\sum_{j,q}\sum_{{\bf i}_q}&\left(-\frac{1}{2 m a^2}\right)^q \e^{-\frac{\beta}{m a^2}} 
\e^{-\beta[V(j_0 a),\ldots,V(j_q a)]}\nonumber\\&\times\delta(j_q,j_0),
\end{align}
where $\delta(j_q,j_0)$ is a Kronecker delta function (with a slight abuse of notation). 
We now show explicitly that in the continuum limit $a\to0$, the walk sum representation of $Z_a$ approaches the Feynman path integral partition function, Eq.~\eqref{eq:FeyZ}. 

To begin, for reasons that will become apparent shortly, let us rewrite $Z_a$ as
\begin{align} 
Z_a=\sum_{j,q}&\left[\sum_{{\bf i}_q}\frac1{2^q}\right] \left[\frac{\left(\frac{\beta }{m a^2}\right)^q}{q!} \e^{-\frac{\beta}{m a^2}} \right]\\\nonumber&\times
 \left[ \frac{q!}{(-\beta)^q} \e^{-\beta[V(j_0 a),\ldots,V(j_q a)]} \right]\delta(j_q,j_0)\,.
\end{align}
Next, we denote $q_\star {\equiv} \frac{\beta }{m a^2}$ and observe that for large $q$ the dependence of $q_\star^q/q!$ on $q$ becomes sharply peaked 
at $q \approx q_\star$, with the rest of the $q$ values  decaying rapidly~\cite{curtis}. Therefore, we can write
\begin{align}
\frac{\left(\frac{\beta}{m a^2}\right)^q}{q!} \e^{-\frac{\beta}{m a^2}} \approx \frac1{\sqrt{2 \pi q_\star}}\e^{-\frac{(q-q_\star)^2}{2q_\star}}\approx \delta(q,q_\star) \,,
\end{align}
and thus,
\begin{align} 
Z_a\approx\sum_j &\left[\sum_{{\bf i}_{q_\star}}\frac1{2^{q_\star}}\right] 
 \left[ \frac{{q_\star}!}{(-\beta)^{q_\star}} \e^{-\beta[V(j_0 a),\ldots,V(j_{q_\star} a)]} \right]\nonumber\\&\times \delta(j_{q_\star},j_0)\,,
\end{align}

\begin{figure}[h!]
\begin{center}
\includegraphics[width=0.99\columnwidth]{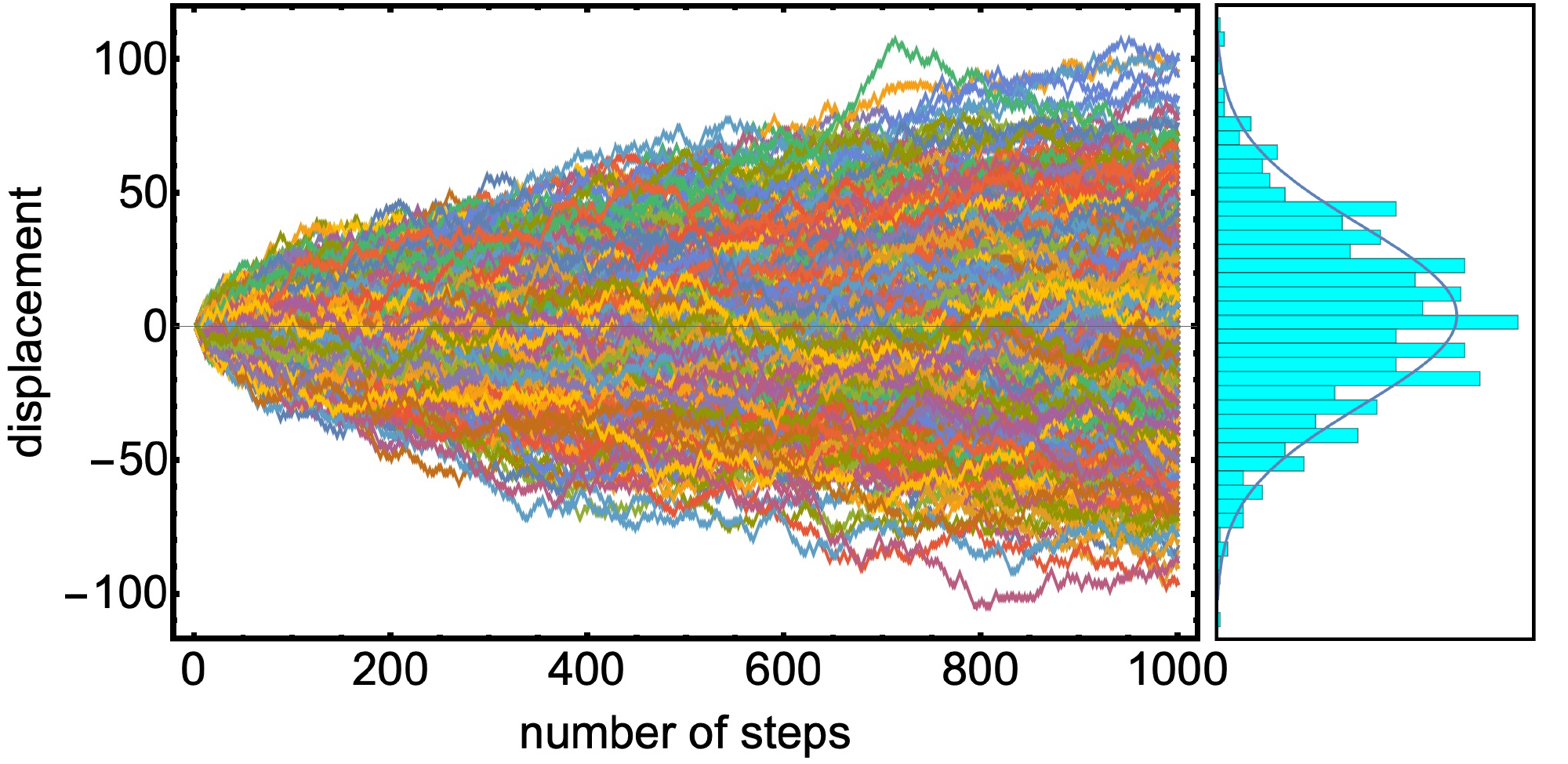}
\end{center}
\caption{{\bf The displacement of a many-step i.i.d. random walk may be approximated by a (continuous) normal variable}. Left: Displacement as a function of number of steps for a thousand i.i.d. random walkers. The total number of steps for each walker is $n=1000$, where each step displaces the walker by one unit either up or down. 
Right: A histogram of the final displacement values of the random walkers depicted in the left panel. In the limit of infinitely-long walks, the distribution approaches a normal one with mean zero and variance $n$.}
\label{fig:normal}
\end{figure}

Next, we notice that we can interpret the term $\sum_{{\bf i}_{q_\star}}\frac1{2^{q_\star}}$ above as a probability measure over all possible walks of $q_\star$ steps. Considering the $q_\star$ steps as $n$ concatenated segments of $N$ steps each ($q_\star=n N$, with $n$ and $N$ chosen such that they are both proportional to $1/a$)
\begin{align}
 \sum_{{\bf i}_{q_\star}}\frac1{2^{q_\star}}  &= \left(\sum_{i_1=\pm 1} \cdots \sum_{i_n=\pm 1} \frac1{2^{n}} \right)\\\nonumber&\times
 \left( \sum_{i_{n+1}=\pm 1} \cdots \sum_{i_{2n}=\pm 1} \frac1{2^{n}}\right)\cdots\\\nonumber&\times
 \left( \sum_{i_{(N-1)n+1}=\pm 1} \cdots \sum_{i_{nN}=\pm 1}  \frac1{2^{n}} \right)\,.
\end{align}
We can view each term above, $\sum_{i_{n (k -1)+1}=\pm 1} \cdots \sum_{i_{nk}=\pm 1}\frac1{2^{n}}$ (with $k=1,\ldots,N$), as a summation over $n$ i.i.d. random steps $i_j=\pm1$. There are $2^{n}$ such $n$-step paths in each term, each of which occurring with probability $2^{-n}$. Equivalently, we can write the sum over $n$ consecutive steps as a sum over all possible displacements defined as $\tilde{s}_k = \sum_{\ell=n (k -1)+1}^{n k}  i_\ell=j_{nk}-j_{n(k-1)}$,  where $\tilde{s}_k$ takes values in the set $s\equiv\{-n,-n+2,\cdots,n-2,n\}$ with probability $\frac1{2^n} {n \choose \frac{\tilde{s}_k}{2}+\frac{n}{2}}$:
\beq
\sum_{i_{n (k -1)+1}=\pm 1} \cdots \sum_{i_{nk}=\pm 1}\frac1{2^n}=\sum_{\tilde{s}_k\in{s}} \frac1{2^n} {n \choose \frac{\tilde{s}_k}{2}+\frac{n}{2}}.
\eeq
At this point it is useful to switch from $\tilde{s}_k$ to a summation variable  $\tilde{\jmath}_k\equiv j_{nk}=j_{n(k-1)}+\tilde{s}_k\in s+j_{n(k-1)}$. The random variable $\tilde{\jmath}_{k}$ represents the dimensionless displacement of the particle with respect to its previous position after $n$ discrete steps. Taking the starting point to be $\tilde{\jmath}_0=0$ for simplicity, we can write
\begin{align}
 \sum_{{\bf i}_{q_\star}}\frac1{2^{q_\star}}  &= \left(\sum_{\tilde{\jmath}_1\in{s}} \frac1{2^n} {n \choose \frac{\tilde{\jmath}_1+n}{2}}\right)\nonumber\\&\times\left(\sum_{\tilde{\jmath}_2\in{s+\tilde{\jmath}_1}} \frac1{2^n} {n \choose \frac{\tilde{\jmath}_2-\tilde{\jmath}_1+n}{2}}\right)\cdots\nonumber\\&\times\left(\sum_{\tilde{\jmath}_N\in{s+\tilde{\jmath}_{N-1}}} \frac1{2^n} {n \choose \frac{\tilde{\jmath}_{N}-\tilde{\jmath}_{N-1}+n}{2}}\right)\,.
\end{align}
Since the expectation value of $i_j$ is zero and its variance is one, for all $j$, it immediately follows from the central limit theorem  that for large  $n$ (we remind the reader that $n$ increases as $1/a$ where $a$ tends to $0$)
\beq
\sum_{\tilde{\jmath}_k\in{s+\tilde{\jmath}_{k-1}}}{\!\!\!\!\!\!} \frac1{2^n} {n \choose \frac{\tilde{\jmath}_{k}-\tilde{\jmath}_{k-1}+n}{2}}{\overset{d}{\longrightarrow}}\frac1{\sqrt{2 \pi n}}\int {\!\!}\rmd \jmath_k \e^{-\frac1{2}\frac{(\jmath_k-\jmath_{k-1})^2}{n}}  
\eeq
where $\jmath_k$ is the continuous counterpart of $\tilde{\jmath}_k$, with $k=1,\ldots,N$, and $\overset{d}{\longrightarrow}$ denotes convergence in distribution. This is illustrated in Fig.~\ref{fig:normal}. 

The integrated probability measure thus converges in distribution to a product of Gaussian distributions:
\begin{align}
 \sum_{{\bf i}_{q_\star}}\frac1{2^{q_\star}}&\overset{d}{\longrightarrow}\left(\frac1{\sqrt{2 \pi n}} \int \rmd \jmath_1 \e^{-\frac1{2}\frac{(\jmath_1-\jmath_0)^2}{n}}\right)\\\nonumber&\times 
\left( \frac1{\sqrt{2 \pi n}}\int \rmd \jmath_2 \e^{-\frac1{2}\frac{(\jmath_2-\jmath_1)^2}{n}}\right) \cdots\\\nonumber&\times
\left(\frac1{\sqrt{2 \pi n}} \int \rmd \jmath_N \e^{-\frac1{2}\frac{(\jmath_{N}-\jmath_{N-1})^2}{n}}\right)\\\nonumber&=\Big(\frac{1}{2\pi n}\Big)^\frac{N}{2} \Big[\prod_{k=1}^N\int \rmd \jmath_k\Big] 
\e^{-\frac1{2} \frac{\sum_{k=1}^{N}(\jmath_{k}-\jmath_{k-1})^2}{n}}.
\end{align}
Switching to the (continuous) dimensional displacement $x_k=\jmath_k a$, we obtain 
\begin{align}\label{eq:measure}
&\sum_{{\bf i}_{q_\star}}\frac1{2^{q_\star}}{\overset{d}{\longrightarrow}} \Big(\frac1{2 \pi n a^2}\Big)^\frac{N}{2}\Big[\prod_{k=1}^N\int \rmd x_k\Big]
\e^{-\frac{\sum_{k=1}^{N}(x_{k}-x_{k-1})^2}{2n a^2}}\nonumber\\
&=\Big(\frac{m}{2 \pi \Delta t}\Big)^\frac{N}{2}\Big[\prod_{k=1}^N\int \rmd x_k\Big]
\e^{-\frac{m}{2} \sum_{k=1}^{N}\big(\frac{x_{k}-x_{k-1}}{\Delta{t}}\big)^2{\!\!}\Delta{t}},
\end{align}
where in the last equation we identified $\Delta{t}=mna^2$. This choice ensures that $N\Delta{t}=q_\star ma^2=\beta$. Upon choosing $n\sim N\sim\frac1{a}$,  in the continuum limit $a\to0$ we have $\Delta{t}\sim a\to 0$ and both $n$ and $N$ tend to $\infty$. Thus in the continuum limit of $a \to 0$, the right-hand-side of Eq.~\eqref{eq:measure}, together with imposing the boundary condition $j_{q_\star}=j_0$ and summing over all starting points $j_0$, recovers the famous Feynman path integral of a free quantum mechanical particle~\cite{PIbook}
\begin{align}\label{eq:FeynmanK}
\lim_{a\to0}\sum_j \sum_{{\bf i}_{q_\star}}\frac1{2^{q_\star}}&\overset{d}{\longrightarrow}\oint{\cal D}x(t)\e^{-\frac{m}{2}\int_{0}^\beta{\dot{x}(t)}^2\rmd{t}} \,.
\end{align}

So far, we have not considered the potential term in $Z_a$. In the Appendix, we  show that in the limit of large $q_\star$ one can write: 
\begin{align}
&\frac{(q_\star)!}{(-\beta)^{q_\star}}  \e^{-\beta[V(j_0 a),\ldots,V(j_{q_\star} a)]} \approx \exp(- \frac{\beta}{q_\star} \sum_{k=0}^{q_\star} V(j_k a))\nonumber\\&\approx\exp(- \frac{\beta}{q_\star} \sum_{k=1}^{q_\star} V(x_k))
\end{align}
where in the absence of a potential this term is identically 1. The right-hand-side of the above equation can be written as
\begin{align}
&e^{- \frac{\beta}{q_\star} \sum_{k=1}^{q_\star} V(x_k)} = \exp\Big(- \frac{\Delta{t}}{n} \sum_{k=1}^{N} \sum_{\ell=1}^{n} V(x_{(k-1)n+\ell})\Big)
 \nonumber\\&=\exp\Big(- \Delta{t} \sum_{k=1}^{N} \frac{\sum_{\ell=1}^{n} V(x_{(k-1)n+\ell})}{n}\Big)\,.
\end{align}
and since in the limit $n\to\infty$, we can replace the average
$\frac{\sum_{\ell=1}^{n} V(x_{(k-1)n+\ell})}{n}$ by $V(x_k)$, in the limit $N \to \infty$, we directly obtain
\begin{align}\label{eq:FeynmanP}
\lim_{a\to0}\exp\left(- \Delta t \sum_{k=1}^{N} V(x_k)\right)=\e^{- \int_0^\beta \rmd t V(x(t))} \,.
\end{align}
Together, Eqs.~\eqref{eq:FeynmanK} and~\eqref{eq:FeynmanP} complete our derivation of the Feynman path integral for the partition function of a one dimensional quantum particle as a continuous limit of the the proposed walk sum of the corresponding discrete quantum system. We have thus demonstrated that the proposed walk sum formalism may serve as a natural extension of the Feynman path integral formalism to discrete-variable quantum systems [see Fig.~\ref{fig:walks}(b)].

Analogously, we can readily derive the Feynman path integral for the transition amplitude, Eq.~\eqref{eq:FeyA}  from its discrete version.   To do that, let us consider without loss of generality the case where the particle is localized at the origin at $t=0$, that is, $x_{\rm in}\equiv x(0)=0$. Following the derivation for the partition function already obtained, one can similarly arrive at the expression  
\begin{align}
\lim_{a\to0}&\bra{x_{\rm out}}e^{-\beta H_a}\ket{x_{\rm in}}\\\nonumber&=\int_{x(0)=x_{\rm in}}^{x(\beta)=x_{\rm out}} {\cal D}x(t) \e^{-\int_0^\beta \rmd t \left(\frac{m}{2 } \dot{x}(t)^2 + V(x(t))\right)} \,.
\end{align}
Applying a Wick rotation $\beta \to i t$~\cite{wick}, we immediately obtain
the desired result for transition amplitudes
\begin{align}
\lim_{a\to0}&\bra{x_{\rm out}}e^{-it H_a}\ket{x_{\rm in}}\\\nonumber&=\int_{x(0)=x_{\rm in}}^{x(t)=x_{\rm out}} {\cal D}x(t) \e^{i\int_0^t\rmd \tau \left(\frac{m}{2 } \dot{x}(\tau)^2 - V(x(\tau))\right)} \,.
\end{align}

\section{Discrete-variable path integrals: an example\label{sec:ising}}


To illustrate the applicability and  potential power of the discrete-variable path integral formulation, we now explicitly work out a transition amplitude calculation for a large-scale  transverse-field Ising model~\cite{Ising} of spin-1/2 particles. The model is crucial in quantum physics as it serves as a fundamental model for studying quantum phase transitions and the interplay between quantum fluctuations and magnetic ordering.

The Hamiltonian of the transverse-field Ising model is given by 
\beq
H= h \sum_i X_i +\sum_{\langle i j \rangle} J_{ij} Z_i Z_j \,,
\eeq 
where $X_i$ and $Z_i$ are Pauli-$X$ and Pauli-$Z$ operators acting on the $i$-th spin, respectively, and ${\langle i j \rangle}$ denotes a summation over the interaction graph of the model (which may be arbitrary). The coefficients $J_{ij}$ correspond to the strengths of the longitudinal interactions between pairs of spins and $h$ is the strength of the external magnetic field acting on the transverse components of the spins.   The Hamiltonian in PMR form is written as $H=D_0+ \sum_i D_i P_i$ where $D_0 = \sum_{\langle i j \rangle} J_{ij} Z_i Z_j$, whereas for $i>0$ the diagonal operators are $D_i=h \cdot \mathds{1}$ and the permutation operators are simply $P_i=X_i$. 

Having cast the Hamiltonian in PMR form, one can now directly write an expression for transition amplitudes between basis states, each of which may be denoted as a sequence of spins pointing either up or down in the $z$ direction, e.g., $| \ua_1 \da_2 \ldots  \da_n\rangle$. For an $n$-spin system there would be $2^n$ such basis states. In the transverse-field Ising model, the action of the permutation operator $P_i=X_i$ on a basis state flips the $i$-th spin:  
\beq
| \dots \ua_i \ldots  \rangle \leftrightarrow  | \dots \da_i \ldots  \rangle \,.
\eeq
Let us now pick two basis states and calculate the transition amplitude from one to the other. For concreteness, let us choose $|z_{\textrm{in}}\rangle = | \ldots \ua \ua \ua \ldots \rangle$ and $|z_{\textrm{out}}\rangle = | \ldots \da \da \da\ldots\rangle$ with all other unspecified spin orientations assumed to be the same between the two states. In this scenario, the discrete-variable counterpart of Feynman paths would be the walks included in the sum
\begin{align}\label{eq:trans}
&\zl\e^{-\frac{i}{\hbar} t H} \zr=\sum_{{\bf i}_q : \zinout} D_{{\bf i}_q}   \e^{-\frac{i}{\hbar} t [E_{z_0},\ldots,E_{z_q}]} \nonumber\\
&= \sum_{{\bf i}_q :  | \ldots \ua \ua \ua \ldots \rangle \to  | \ldots \da \da \da \ldots \rangle} h^q   \e^{-\frac{i}{\hbar} t [E_{(\ldots \ua \ua \ua \ldots )},\ldots,E_{(\ldots \da \da \da \ldots )}]}  
\,, 
\end{align}
where ${\bf i}_q :  | \ldots \ua \ua \ua \ldots \rangle \to  | \ldots \da \da \da \ldots \rangle$ denotes all sequences of length $q$ of single spins flips that take the state $|z_{\textrm{in}}\rangle=| \ldots \ua \ua \ua \ldots \rangle $ to 
$|z_{\textrm{out}}\rangle=| \ldots \da \da \da \ldots \rangle$ and where the diagonal energies, which are the inputs to the divided difference exponential are defined as 
\beq
E_{(\ldots \ua \ua \ua \ldots )}=  \langle \ldots \ua \ua \ua \ldots | D_0 | \ldots \ua \ua \ua \ldots \rangle \,,
\eeq
with analogous expressions for all other basis states. 
\begin{figure}[htp]
\includegraphics[width=.9\columnwidth]{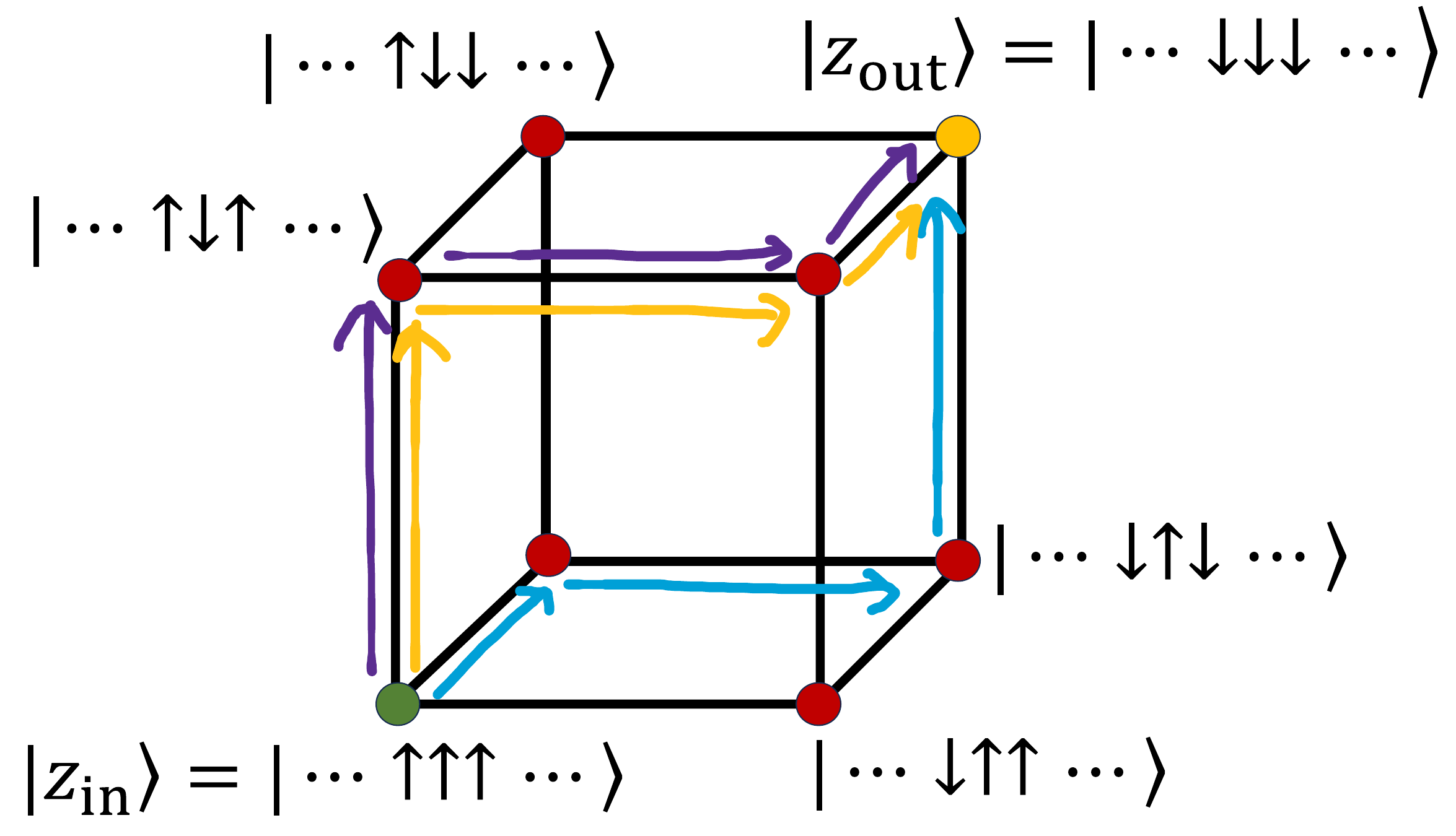}
\caption{\label{fig:qmcWeights} {\bf A diagrammatic representation of walks contributing to a transition amplitude between $|z_{\textrm{in}}\rangle =|\cdots \ua \ua \ua \cdots \rangle$ and $|z_{\textrm{out}}\rangle =|\cdots \da \da \da \cdots \rangle$ in the transverse-field Ising model}. Since the in and out states are three spin-flips apart, the shortest walks contributing to the transition amplitude are the six possible paths on the hypercube of the basis states corresponding to the six different orderings in which the three spins may be flipped. Three of the six paths are shown in the figure as sequences of arrows. Each arrow, which points from one node to its neighbor, corresponds to a single spin flip, i.e., the action of a single permutation operator, namely a Pauli-$X$, on a basis state. }
\end{figure}
In Fig.~\ref{fig:qmcWeights} we provide a diagrammatic illustration of the above, where for example, the contribution from the specific walk $|\ua\ua\ua\rangle \to |\ua\ua\da\rangle \to |\ua\da\da\rangle \to |\da\da\da\rangle$ evaluates to 
 $h^3 \e^{-i \frac{t}{\hbar} [E_{\ua\ua\ua},E_{\ua\ua\da},E_{\ua\da\da},E_{\da\da\da} ]}$ with similar expressions for all other walks. The sum total of all contributions from all walks connecting the two end states is the total transition amplitude.  

Aside from transition amplitudes, one may also express the partition function $Z=\tr \e^{-\beta H}$ as a sum of contributions from various walks in much the same way, except that for $Z$ there is a weight associated with every closed walk on the Hamiltonian graph and the sum of those constitutes $Z$. Since the partition function encodes all the thermodynamic properties of a system in equilibrium, summing over or sampling these partition function terms allows one to derive key quantities for various quantum spin (or other discrete-variable) systems including free energy, internal energy, entropy, specific heat and fidelity susceptibility. The calculation of those and other quantities may lead in turn to a better understanding of phase transitions, critical phenomena, the distribution of particles in different energy states and more, taking place in the tested model (see, e.g., Refs.~\cite{pmr,PhysRevResearch.6.013281,ezzell2024exactuniversalquantummonte}).

\section{Summary and conclusions}

We proposed, for the first time to the authors' knowledge, a universal, parameter-free discrete-variable analog to Feynman's path integral formulation. We demonstrated that for discrete systems path integrals take the form of walks on the Hamiltonian graph. We  further showed that Feynman path integrals can in fact be derived from the proposed discrete analog in the continuum limit. 
The methods introduced here also offer a universal discretization scheme for continuous-variable quantum systems.  

Feynman's path integral formulation has been transformative in advancing our understanding of both equilibrium and non-equilibrium phenomena in many-body quantum systems with a focus on dynamic correlations and emergent properties~\cite{feynman1,feynman2}. We hope that formulation presented here, aside from being a complement to Feynman's representation of quantum mechanics, will be leveraged in the future to shed new light on quantum phenomena in much the same way as Feynman's original integrals are used to obtain important insights about continuous-variable quantum systems, specifically, the temporal evolution of spin correlations in geometrically frustrated lattices~\cite{Miz06Gapless}, quantum phase transitions in discrete spin models, such as the transverse-field Ising model~\cite{Tak07Quantum} and beyond. 

One direct application of walk sums is that they lend themselves easily to numerical methods. Quantum Monte Carlo simulations of quantum many-body systems~\cite{pmr,PhysRevResearch.6.013281,PhysRevB.109.134519} and other computational techniques~\cite{fmat} based on the idea of walk sums have already proved themselves useful for studying the properties of complex large-scale quantum many-body systems. It is worth noting that in the context of quantum Monte Carlo, one may view the walk sums approach presented here as analogous to continuous-time path integral quantum Monte Carlo~\cite{qpt:book,PhysRevE.74.036701,PROKOFEV1998253}, with the critical difference that the divided-difference weights of the walk sums technique represent fully integrated continuous-time path integral Monte Carlo paths owing to the divided differences formulation. This is in contrast with the continuous-time path integral procedure in which the multi-dimensional time integrals are sampled. 

\begin{acknowledgments}
AK acknowledges support for this project by NSF award \#2210374. 
IH acknowledges support by the Office of Advanced Scientific Computing Research of the U.S. Department of Energy under Contract No DE-SC0024389. 
\end{acknowledgments}

\bibliography{refs}

%
\section*{Appendix: The divided difference exponential in the large-number-of-inputs limit}
Consider a (multi-)set of $q+1$ real-valued inputs $x_0,\ldots,x_q$ with mean $\mu = (\sum_{i=0}^q x_i)/(q+1)$ and variance $\sigma^2 = \sum_{i=0}^q (x_i - \mu)^2/(q+1)$. 
We first prove\footnote{A special thanks is extended to Lev Barash for the main proof detailed in this appendix.} that for $q \gg \sigma^2$ the following holds.
\beq
q!\e^{[x_0,\dots,x_q]} = \e^{\left(\mu+ \frac{\sigma^2}{2(q+2)}+O\left(\frac{1}{q^2}\right)\right)} \,. 
\label{eq:ddinfty}
\eeq
To do that, we use the Hermite-Genocchi formulation of divided differences~\cite{baxter2010functionals} 
\beq
f{[x_0,\ldots,x_q]}= \int_\Omega  f^{(q)}\left(\sum_{i=0}^q t_i x_i\right)  \rmd t_1 \rmd t_2 \dots \rmd t_q\,,
\eeq
where $f(\cdot)$ is arbitrary Taylor-expandable function,  $f^{(q)}(\cdot)$ denotes the $q$-th derivate of $f(\cdot)$, the volume of integration $\Omega$ is defined by $t_i>0$ for all $i$, $\sum_{i=1}^q t_i<1$,
and $t_0 = 1 - \sum_{i=1}^q t_i$. 
For the exponential function the above reads
\beq
\e^{[x_0,\ldots,x_q]}= \int_\Omega  \e^{\sum_{i=0}^q t_i x_i}  \rmd t_1 \rmd t_2 \dots \rmd t_q\,,
\eeq

First, we note that $\e^{[x_0-\mu,\dots,x_q-\mu]}= \e^{-\mu}\e^{[x_0,\dots,x_q]}$ which can be shown directly from the Hermite-Genocchi formulation
\bea
\e^{[x_0,\ldots,x_q]}&=& \e^\mu \int_\Sigma  \e^{\sum_{i=0}^q t_i (x_i-\mu)}  \rmd t_1 \rmd t_2 \dots \rmd t_q
\nonumber \\&=& \e^{[x_0-\mu,\ldots,x_q-\mu]} \,,
\eea
where we have used the fact that $\sum_{i=0}^q t_i =1$. 
Due to the above, we may assume without loss of generality that $\mu = 0$ for what follows.

Expanding the integrand in the Hermite-Genocchi formulation in a Taylor series, we find that
\beq
\e^{[x_0,\ldots,x_q]} =
\int_\Omega 
\left( 1 + s_q + \frac{s_q^2}{2} + \ldots \right) \rmd t_1 \rmd t_2 \dots \rmd t_q,
\label{eq:HGexpansion}
\eeq
where 
$s_q = \sum_{i=0}^{q} t_i x_i$. In addition, it is easy to work out
\begin{align}
&\int_\Omega \left(\sum_{i=0}^q x_i t_i\right)^m  \rmd t_1 \rmd t_2 \dots \rmd t_q \\\nonumber&= 
\frac{m!}{(q+m)!} \sum_{\substack{k_0+\ldots+k_q = m \\ 0 \leq k_j \leq m }} \,\,\prod_{j=0}^q x_j^{k_j},
\label{eq:L4}
\end{align}
where $m\geq 0$. Specifically:
\ba
\int_\Omega \rmd t_1 \rmd t_2 \dots \rmd t_q &=& \frac{1}{q!}, \\
\int_\Omega s_q \rmd t_1 \rmd t_2 \dots \rmd t_q &=& \frac{\mu}{n!} = 0,\\
\int_\Omega \frac{s_q^2}{2} \rmd t_1 \rmd t_2 \dots \rmd t_q &=& \frac{1}{2(q+2)!} \Bigg(\!\!\! {\Big(\sum_{i=0}^q x_i \Big)^2 + \sum_{i=0}^n x_i^2}\!\Bigg)\\
\int_\Omega \frac{s_q^3}{6} \rmd t_1 \rmd t_2 \dots \rmd t_q &=& \frac{1}{(q+3)!}
\Bigg(
\sum_i x_i^3 \\\nonumber&+& \sum_{i < j} ( x_i^2 x_j + x_i x_j^2) + \sum_{i < j < k} x_i x_j x_k
\Bigg).
\ea
Hence, it follows from Eq.~(\ref{eq:HGexpansion}) that
\beq
q! \e^{[x_0-\mu,\ldots,x_q-\mu]}= 1 + \frac{\sigma^2}{2(q+2)} + O\left(\frac{1}{q^2}\right) \,,
\eeq
or, expressed differently
\beq
q! \e^{[x_0,\ldots,x_q]} \approx \e^\mu \left( 1 + \frac{\sigma^2}{2(q+2)}\right)\,.
\eeq
Next, from the Hermite-Genocchi formulation for $f(\cdot)=\e^{-\beta(\cdot)}$, one can easily show that
\beq
\e^{[-\beta x_0,\ldots,-\beta x_q]} = \frac{1}{(-\beta)^q} \e^{-\beta[x_0,\ldots,x_q]} 
\eeq
Putting the above together, we find that for any fixed variance $\sigma$, in the limit of $q \to \infty$, we get
\begin{align}
q! \e^{[-\beta x_0,\ldots,-\beta x_q]} &= \frac{q!}{(-\beta)^q} \e^{-\beta[x_0,\ldots,x_q]} \\\nonumber& \approx \e^\mu =\e^{-\frac{\beta}{q+1} \sum_{j=0}^q x_j}\,,
\end{align}
 as asserted in the main text. \vfill
\end{document}